# Higgs physics in superconductors[*][†]

Hao Chu[‡], Haotian Zhang, Zhili Zhang

(Shanghai Jiao Tong University, School of Physics and Astronomy, Shanghai 200240)

**Abstract**

As pointed out by Nambu-Goldstone theorem, the breaking of continuous symmetry gives rise to massless or gapless bosonic excitations. In superconductors, continuous local U(1) gauge symmetry is broken. The gapless excitation thus created is the collective phase mode of the superconducting order parameter. In 1962, Philip Anderson pointed out that the Coulomb interaction between Cooper pairs lifts this gapless mode to the superconducting plasma frequency. Therefore, in a superconducting fluid there are no bosonic excitations below the binding energy of the Cooper pairs ($2\Delta$). Anderson's mechanism also implies that the massless photon, which mediates electromagnetic interaction, becomes massive in a superconductor. This mechanism provides a microscopic theory for the dissipationless charge transport (in conjunction with Landau's criterion for superfluidity) as well as the Meissner effect inside a superconductor. Jumping to particle physics, in order to explain why the gauge bosons for electroweak interaction, namely the W±, Z bosons, are massive, in 1964 Peter Higgs, François Englert, Tom Kibble and colleagues proposed the existence of a field (presently referred to as the Higgs field) in nature. This matter field couples to the massless W±, Z bosons and generates mass via the Higgs mechanism. Due to their conceptual similarities, these two mechanisms are collectively referred to as the Anderson-Higgs mechanism. In 2013, the detection of the scalar excitation of the Higgs field, namely the Higgs boson, at the Large Hadron Collider provided the final proof for the Higgs hypothesis almost 50 years after its conception. The amplitude mode of the superconducting order parameter, which corresponds to the Higgs boson through the above analogy, is referred to as the Higgs mode of a superconductor. Its spectroscopic detection has also remained elusive for nearly half a century. In recent years, the development of ultrafast and nonlinear spectroscopic techniques enabled an effective approach for studying the Higgs mode in superconductors. In this review, we introduce the Higgs mode from a perspective of why superconductors superconduct and review the recent development in Higgs spectroscopy, particularly in the field of ultrafast and nonlinear terahertz spectroscopy, for non-expert readers. We then discuss the novel perspectives and insights that may be learnt from these studies for future high-temperature superconductivity and correlated materials research in general.

---

[*] The paper is an English translation of the manuscript originally published in Chinese in Acta Physica Sinica. Please cite the paper as: **CHU Hao, ZHANG Haotian, ZHANG Zhili, Higgs physics in superconductors.** *Acta Phys. Sin.*, 2025, (11): 117402. doi: <u>10.7498/aps.74.20250241</u>

[†] Project supported by the National Natural Science Foundation of China (Grants No. 12274286), the National Key Research and Development Program of China (Grants No. 2024YFA1408701) and the Yangyang Development Fund.

[‡] Corresponding author: haochusjtu@sjtu.edu.cn

**Anderson-Higgs mechanism**

In 1957, Bardeen, Cooper and Schrieffer proposed that phonon-mediated pairing of $k_\uparrow$ and $-k_\downarrow$ electrons on the Fermi surface yields a lower-energy bound state later known as the Cooper pair [1,2], providing a microscopic theory for conventional superconductivity. In the half century that followed, BCS theory achieved great success. Accurate predictions were made for the superconducting transition temperature ($T_c$) of various conventional superconductors as well as the highest possible $T_c$ of phonon-mediated superconductors [3]. The success of the BCS theory, coupled with the long-standing mystery of the pairing mechanism behind high-temperature superconductivity, makes electron-pairing almost synonymous with superconductivity itself. When explaining the phenomenon of the critical velocity $v_c$ of superfluid helium (i.e. superfluid helium loses superfluidity if it flows at a velocity greater than $v_c$), Landau pointed out that the interaction between bosons, which transforms their single-particle-like excitations into collective excitations, holds the key for dissipationless transport [4]. For instance, consider a Bose-Einstein condensate of atoms, in which all atoms reside in their $q = 0$ energy ground state. Imagine that the condensate is residing in a container that moves at a constant velocity $v$ relative to the superfluid. If we perform Lorentz transformation on the condensate to the container's rest frame, the excitation spectrum of the condensate is shifted due to the Doppler effect. When $|v|$ is large enough such that the $q \neq 0$ excited states of the cold atoms are Doppler-shifted to zero energy, then thermal fluctuations from the container or within the condensate will cause spontaneous excitations of the atoms from the $q = 0$ ground state. The Bose-Einstein condensate is thus broken and dissipationless transport is lost. Specifically, for non-interacting bosons characterized by a free particle-like dispersion, their condensate cannot sustain any nonzero $v_c$ (Fig. 1(a)). A superfluid with nonzero $v_c$ can only emerge from systems where interaction modifies the excitation spectrum into non-free-particle-like. In superfluid helium-4, the van der Waals interaction between the helium atoms leads to phonon-like collective modes [5]. In this case, $v_c$ corresponds to the phase velocity for spontaneously generating roton excitations (Fig. 1(b)). Similarly, in superconductors the critical current $I_c$ ($\propto 2en_s v_c$, where $n_s$ is the superfluid density) corresponds to spontaneous excitations inside the Cooper-pair condensate when it travels at a velocity $v_c$ relative to the lattice. This implies that in order to improve the design of superconducting devices or discover new kinds of superconductors with better performance, a close understanding of the collective response of superconductors is indispensable.

To describe the collective modes of a superconductor, first we point out that the superconducting transition breaks local U(1) gauge symmetry. Specifically, above $T_c$ the free energy of a superconductor is invariant with respect to local rotations of the phase $\phi(x)$ of the superconducting order parameter $\psi(x) = |\psi(x)|e^{i\phi(x)}$. Therefore, the local $\phi(x)$ can be regarded as a redundant degree of freedom, which in fact corresponds to the gauge freedom in electromagnetism as will be discussed below. During condensation, Cooper pairs globally develop a coherent phase, which is also the phase of the macroscopic superconducting wavefunction. This means that in the superconducting ground state below $T_c$, local rotations of $\phi(x) \in [0, 2\pi]$ cost energy, as also implied by the $K\nabla\psi^*\nabla\psi$ term in the Ginzburg-Landau free energy expression (Eqn. (1)). Such a term implies that local (i.e. short-wavelength) twisting or fluctuation in the order parameter phase leads to an increase in the free energy of the system. However, it is important to note that a superconducting state retains the global U(1) gauge symmetry. Nambu-Goldstone theory states that continuous symmetry breaking leads to the emergence of a new gapless mode.

In a superconductor, such a gapless mode corresponds to the collective phase fluctuation of the superconducting order parameter in the long-wavelength limit. Figure 2(a) shows the free energy surface of a superconductor as a function of its order parameter below $T_c$, with the shape of a Mexican hat. Superconducting phase fluctuations correspond to the order parameter fluctuations along the circumferential direction. The dynamics along the radial direction describe the amplitude fluctuations, or fluctuations in the Cooper-pair binding energy. The curvature of the free energy surface implies that the phase fluctuation is gapless while the amplitude fluctuation is gapped. Their dispersion relations are schematically depicted in Fig. 2(b)[6].

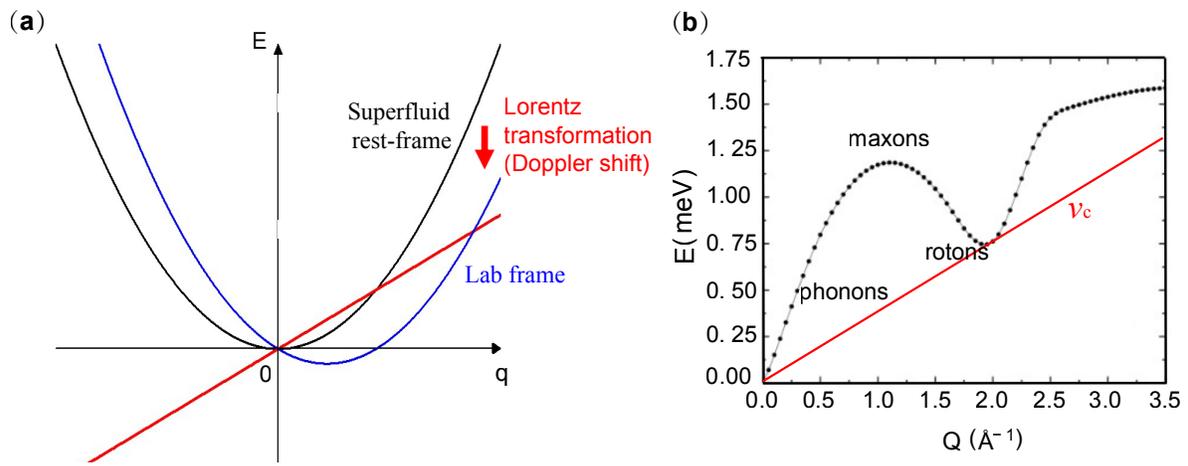

**Figure 1 Landau criteria for superfluidity** (reproduced with permission from Ref. [5]). **(a)** Energy-momentum relation of a free particle. When a particle travels at velocity $v$ (given by the gradient of the red solid line) with respect to its environment (e.g. container, lattice), its excitation spectrum (black solid line) is Doppler-shifted (i.e. to the blue solid line): excitations with momentum/wavevector parallel to $v$ are red-shifted while those anti-parallel to $v$ are blue-shifted. If a state is red-shifted close enough to zero energy, thermal fluctuations between the environment and the particle may cause the particle to be spontaneously excited to that state. The particle no longer remains in its original state, thereby breaking out of the Bose-Einstein condensate and losing superfluidity. **(b)** The van der Waals interaction between the helium atoms causes liquid helium to exhibit phonon-like collective excitations. In the long-wavelength limit (Q → 0), these collective modes exhibit a linear dispersion. As long as superfluid helium flows at a speed smaller than the phase velocity of such acoustic mode, the latter will not be spontaneously excited. The critical velocity of superfluid helium is given by the phase velocity of the roton excitation.

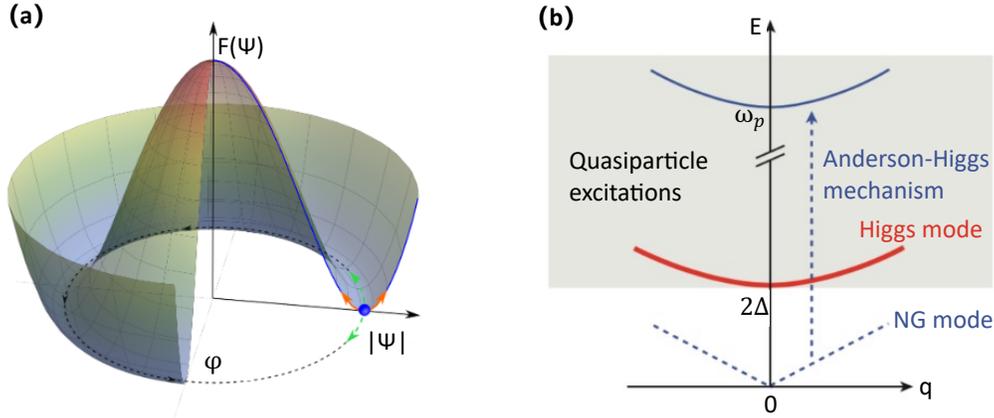

**Figure 2 Collective modes of a superconductor** (reproduced with permission from Ref. [6]). **(a)** Below the superconducting transition temperature $T_c$, the free energy of a superconductor exhibits a Mexican hat-like potential surface with respect to the superconducting order parameter $\psi = |\psi|e^{i\phi}$. The continuous rotational symmetry of the order parameter along the circumferential direction is spontaneously broken. The dynamics along this direction describe the order parameter phase fluctuation; the dynamics along the radial direction describe the order parameter amplitude fluctuation. They are characterized by zero and non-zero frequency, and therefore can be described as massless and massive quasiparticles. **(b)** Dispersions of the phase mode (NG mode) and the amplitude mode (Higgs mode) of the superconducting order parameter [6]. In a superconductor, the massless phase mode becomes massive due to Anderson's mechanism. Its frequency is lifted to the superconductor plasma frequency $\omega_p$. The massive amplitude mode has an energy gap of $2\Delta$ in the long-wavelength limit [12].

$$F = K\nabla\psi^*\nabla\psi + a\psi^*\psi + b(\psi^*\psi)^2 \qquad (1)$$

Given the dispersion of the two collective modes, is the critical velocity of a superconductor given by the phase velocity of the phase mode? In 1962, Philip Anderson pointed out that Coulomb interaction between Cooper pairs lifts this gapless mode to the superconductor plasma frequency [7], ensuring that there are no bosonic excitations below $2\Delta$ within a superconducting fluid. Therefore, $v_c$ is no longer limited by the slope of the phase mode. In the same process, the massless photon which mediates Coulomb interaction also becomes massive. Anderson's mechanism can be simply illustrated using the Ginzburg-Landau theory. To do so, we introduce local amplitude and phase fluctuations in the superconducting order parameter: $\psi_0 \to |\psi_0 + \delta\psi(\vec{x})|e^{i(\phi_0+\delta\phi(\vec{x}))}$, where $|\psi_0|$ and $\phi_0$ denote the global mean-field value of the order parameter amplitude and phase. Substituting this into Eqn. (1) and simplify, we obtain

$$F = F_0 + K\{\nabla\delta\psi(\vec{x})\}^2 + 2|a|\{\delta\psi(\vec{x})\}^2 + K|\psi_0|^2\{\nabla\delta\phi(\vec{x})\}^2 \quad (2)$$

where $F_0$ include all the constant terms. We recognize in Eqn. (2) that the terms containing the amplitude fluctuation $\delta\psi(\vec{x})$ give the equation of motion for massive particles. The single term containing the phase fluctuation $\delta\phi(\vec{x})$ gives the equation of motion for massless or gapless particles, as expected. At this step, we consider the Coulomb interaction between Cooper pairs via minimal coupling (i.e. assuming that electromagnetic field couples only to the electrical charge of a Cooper pair by modifying its momentum). This is done by replacing the gradient operator $\nabla$ with the canonical momentum operator $-i\hbar\nabla + 2e\vec{A}$ (remember that a Cooper pair has charge $2e$). Equation (2) then becomes

$$F = F_0 + K|-i\hbar\nabla\delta\psi(\vec{x})|^2 + 2|a||\delta\psi(\vec{x})|^2 + K|\psi_0|^2\{(-i\hbar\nabla\delta\phi(\vec{x}) + 2e\vec{A})\}^2 + \frac{1}{8\pi}(\nabla\times\vec{A})^2 \quad (3)$$

where the energy density of the electromagnetic field is also added.

To derive Anderson's mechanism, here we perform a crucial step: the local superconducting phase fluctuation $\delta\phi(\vec{x})$ can be absorbed into the vector potential $\vec{A}$ as its gauge without affecting the free energy of the system or any physical observables (recall our earlier statement that the local phase of the superconducting order parameter, a redundant degree of freedom, corresponds to the gauge freedom for describing electromagnetic interaction):

$$\vec{A'} = \vec{A} - i\frac{\hbar}{2e}\nabla\delta\phi(\vec{x}) \quad (4)$$

Using this newly-defined vector potential $\vec{A'}$, Eqn. (3) simplifies into

$$F = F_0 + K|-i\hbar\nabla\delta\psi(\vec{x})|^2 + 2|a||\delta\psi(\vec{x})|^2 + 4e^2K|\psi_0|^2\{\vec{A'}\}^2 + \frac{1}{8\pi}(\nabla\times\vec{A'})^2 \quad (5)$$

Comparing with Eqn. (2), it can be seen that local phase fluctuations of the superconducting order parameter have vanished from the free energy expression. Instead, a massive electromagnetic vector potential remains. This expression can be interpreted as the hybridization of the superconducting phase fluctuation and the electromagnetic field. In other words, $\vec{A'}$ represents a massive electromagnetic field (photon) and a gapped phase mode at the same time, with its mass or energy given by $\sqrt{4e^2K|\psi_0|^2}$. This implies that in mean-field superconductors, the energy of the phase mode (i.e. the superconductor plasma frequency $\omega_p$) is closely related to the superfluid density $n_s = |\psi_0|^2$. In fact, in BCS superconductors the two quantities are inter-defined: $\omega_p = \sqrt{n_s}$ [8]. Empirically, in conventional superconductors the superfluid density is obtained by measuring the London penetration depth which also gives the phase rigidity or phase stiffness of the superfluid. Finally and very importantly, in vacuum the electromagnetic interaction is long-ranged as it is mediated by massless photons. Anderson's mechanism implies that in a superconductor, electromagnetic interaction becomes short-ranged (i.e. exponentially decaying) as a result of the gauge bosons becoming massive. In this sense, Anderson's mechanism also provides a microscopic explanation for the Meissner effect.

Before we end this section, we briefly discuss the analogy between the Anderson mechanism and the Higgs mechanism [9-11]. In electroweak theory, the weak interaction is mediated by W±, Z bosons, in analogy to the electromagnetic interaction mediated by photons. In this analogy, the Higgs field corresponds to the superconducting order parameter. The Higgs field has SU(2) gauge symmetry and is collectively described by four degrees of freedom. Local fluctuations in three of them can be absorbed into the W±, Z bosons by their gauge freedom, thereby making the W±, Z bosons massive. This is similar to the superconductor phase fluctuation being absorbed into photons by their gauge freedom, making the photons massive. The scalar excitation in the remaining degree of freedom of the Higgs field is known as the Higgs boson. It corresponds to the amplitude fluctuation of the superconducting order parameter. Higgs boson is the only independent, detectable degree of freedom of the Higgs field. Therefore, finding its presence within the theoretically-predicted parameter space confirms the validity of the Higgs hypothesis. It is also worth mentioning that in the Standard Model of particle physics, other types of massless particles (e.g. quarks, leptons) also interact with the Higgs field to acquire mass, however, through a different mechanism known as the Yukawa interaction. In a similar sense, in superconductors it is conceivable that the superconducting order parameter interacts with other order parameters or degrees of freedom, affecting their dynamics or ground state properties. Decades after the conception of the Nambu-Goldstone theorem and the Anderson-Higgs mechanism, the close analogy between superconductivity and particle physics continues to manifest and inspire developments in the two fields.

**Spectroscopy Methods for Detecting Higgs Mode**

Unlike the half-century quest for verifying the Higgs mechanism, Meissner effect can be taken as a proof of Anderson's mechanism from the very beginning. Nevertheless, observing signatures of the Higgs mode in superconductors still has meaningful implications. However, experimental search for the Higgs mode faces considerable challenges similar to the Higgs boson. This is partly because Higgs mode describes collective fluctuations in the binding energy of the ($k_\uparrow, -k_\downarrow$) electron pairs. The mode is Raman-active and cannot linearly couple to light in spectroscopy experiments. More importantly, Higgs mode is degenerate with quasiparticle excitation or pair-breaking onsetting at $2\Delta$ energy [12] (Fig. 2(b)). Therefore, any finite-momentum excitation of the Higgs mode would rapidly decay into quasiparticle excitations around the Fermi surface, resulting in an extremely short lifetime of the mode [13,14]. In fact, in Raman scattering studies of superconductors, an inelastic peak centered around $2\Delta$ is often reported. It has been long interpreted as due to pair-breaking peak than the Higgs mode. In addition, the orthogonality between the Higgs mode and the superconducting phase mode is also questioned by theory in certain types of superconductors [15,16], inviting further inquiry about the smoking-gun signature of Higgs mode in spectroscopy. For such reasons perhaps, Higgs mode has remained elusive from spectroscopy studies decades after the proposal of Anderson's mechanism.

In 1980, Sooryakumar and Klein reported Raman signals of the superconducting-charge density wave (CDW) material $2H$-NbSe$_2$ inside a magnetic field [17,18]. During the cooling process, they observe the appearance of two collective modes corresponding to the two states. In the coexisting superconducting-CDW state around 2 K, the application of an external magnetic field suppresses the superconductor mode and enhances the CDW mode. This implies that the two collective modes are coupled (Fig. 3(a)). In the following year, Littlewood and Varma proposed that the collective mode associated with the superconducting state is indeed the long-sought-after Higgs mode [19,20]. In particular, due to the interaction between the superconducting state and the CDW state, the CDW amplitude mode hybridizes with the Higgs mode and renormalizes its energy below the pair-breaking energy $2\Delta$. As such, the Higgs mode cannot decay to quasiparticle excitations, thereby acquiring a long lifetime. In addition, the CDW amplitude mode in theory arises from the folding of an acoustic phonon to the Brillouin zone center due to the CDW ordering.

That is, the CDW amplitude mode is of phonon character and therefore the Higgs mode acquires Raman scattering cross section from the phonon.

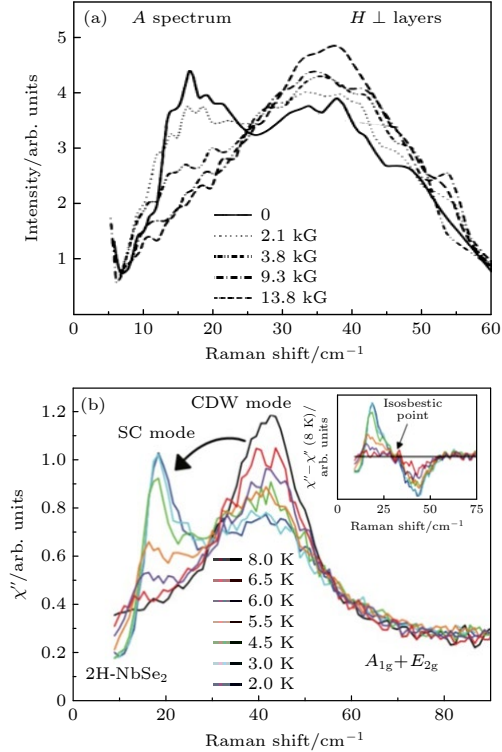

**Figure 3** Raman scattering experiments on 2$H$-NbSe$_2$ (reproduced with permission from Ref. [17] & [55]). **(a)** In the experiment by Sooryakumar and Klein [17], the superconductor collective mode (~ 20 cm$^{-1}$) and the charge density wave (CDW) collective mode (~ 40 cm$^{-1}$) exchange spectral weight under the application of an external magnetic field that suppresses superconductivity. **(b)** By varying temperature, Méasson *et al.* also observe a spectral weight transfer between the superconductor collective mode and the CDW collective mode [55]. Around 6 K, the two modes have approximately equal spectral weight.

Although 2$H$-NbSe$_2$ demonstrated a precedence of spectroscopic detection of the Higgs mode, further reports of Higgs mode in other superconductors remained in standstill for the next 30 years. In 2013, Matsunaga *et al.* reported time-domain free oscillations of the Higgs mode in a Nb$_{1-x}$Ti$_x$N superconducting thin film using terahertz pump-probe technique [21]. In their experiment, a high-field monocycle terahertz pulse generated from LiNbO$_3$ crystal is used to perturb the ground state of the superconductor; meanwhile, another weak terahertz pulse probes the transmissivity of the sample at a variable time delay. By scanning the time delay, the ultrafast evolution of the sample's low-energy transmissivity is obtained, which exhibits a periodic oscillation with a frequency of $2\Delta$ after pump excitation. The oscillation is consistent with the characteristic energy of the Higgs mode (Fig. 4). In 2014, Matsunaga et al. used a multicycle terahertz pulse (of frequency of $\omega$) to drive a NbN superconducting thin film. Below $T_c$, they observed a driven oscillation of frequency $2\omega$ in the terahertz transmissivity of the sample [22] (Fig. 5). This $2\omega$ oscillation is interpreted as the driven collective oscillation of the superfluid, as the superconducting order parameter couples quadratically to electromagnetic field (see Eqn. 5). In the same study, the probe pulse is then blocked and only the pump pulse is let through the sample. The terahertz transmission is found to contain a significant amount of third harmonic generation (THG) below $T_c$ (Fig. 6). By varying temperature $T$ such that $2\Delta(T)$ satisfies $2\Delta(T) = 2\omega$, the THG intensity is found to exhibit a maximum due to the resonance of the Higgs mode under periodic drive.

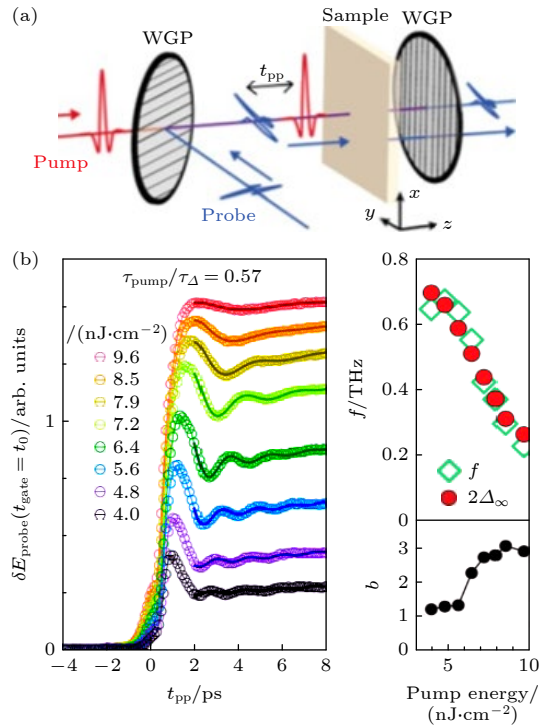

**Figure 4 Time-domain free oscillations of the Higgs mode** (reproduced with permission from Ref. [21]). **(a)** Using a monocycle terahertz pump-probe set-up, Matsunaga et al. studied the ultrafast terahertz transmission of Nb$_{1-x}$Ti$_x$N superconducting thin films. **(b)** After the superconducting ground state is perturbed by a monocycle pump pulse, the terahertz transmissivity of the sample is probed and is found to exhibit a free oscillation with a frequency $f \sim 2\Delta$. With increasing pump pulse energy density, the oscillation frequency $f$ decreases, accompanied also by an increase in the damping constant $b$.

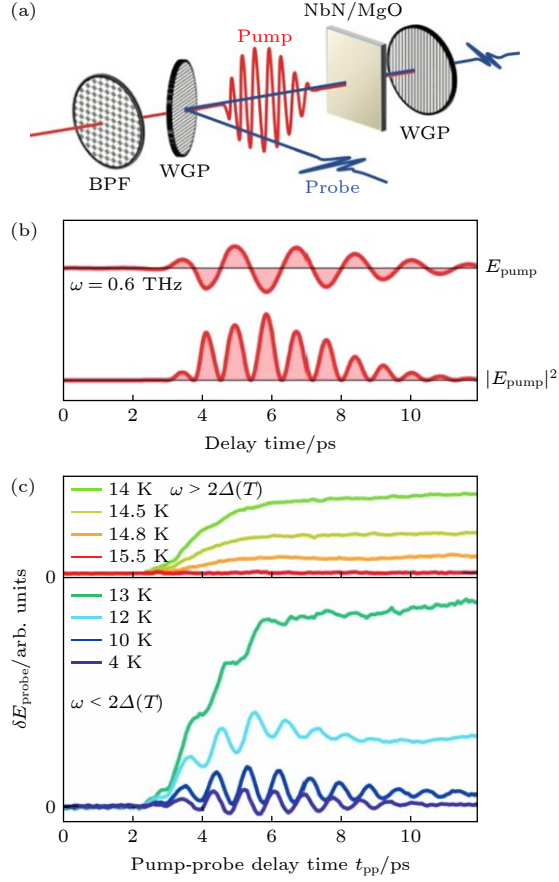

**Figure 5 Higgs oscillation driven by multicycle terahertz pulses** (reproduced with permission from Ref. [22]). **(a)** Multicycle terahertz pump and monocycle terahertz probe set-up. **(b)** The waveform of the multicycle terahertz pulse and its modulus square. **(c)** Under the periodic drive of a multicycle terahertz pulse, the terahertz transmissivity of the sample exhibits characteristic time evolutions above and below $T_c$. Above $T_c$, the terahertz pump induces only thermal heating of the sample, manifested as a monotonic time-evolution of the terahertz transmissivity; below $T_c$, the terahertz pump induces a coherent oscillation of the superfluid at twice the driving frequency, originating from the driven Higgs oscillation.

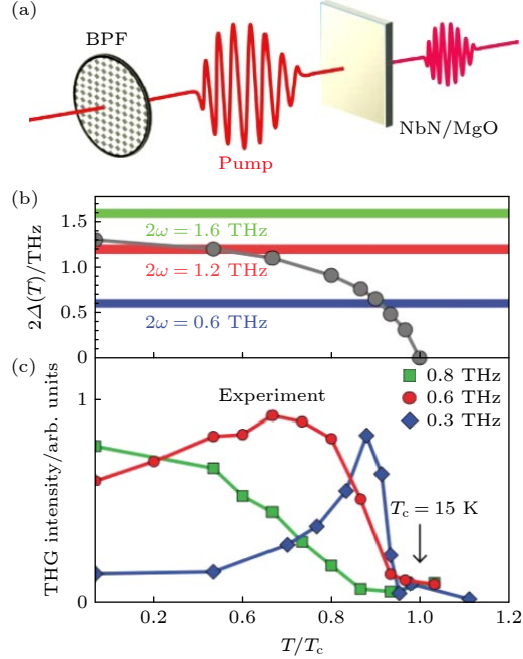

**Figure 6 Terahertz third harmonic generation (THG) from NbN superconductors** (reproduced with permission from Ref. [22]). **(a)** Using a multicycle terahertz pulse to drive the NbN superconducting thin film, Matsunaga et al. observed a clear terahertz THG response from the sample below $T_c$. **(b)** The temperature-dependent superconducting order parameter $2\Delta(T)$ compared to twice the terahertz driving frequency ($\omega$ = 0.3, 0.5, 0.7 THz). **(c)** The temperature dependence of THG intensity under different driving frequencies. At temperatures where $2\Delta(T) = 2\omega$, the driven Higgs oscillation becomes resonant with the driving field, leading to a divergence of the THG intensity.

In the above experiment, the terahertz field-driven dynamics of the superconducting order parameter can be modeled using the Anderson pseudospin model [22-24]. Anderson pseudospin describes the occupation of a pair of electron states with opposite wavevectors and spins ($\boldsymbol{k}_\uparrow, -\boldsymbol{k}_\downarrow$). Pseudospin = 1/2 represents the electron-pair state being occupied, while pseudospin = − 1/2 represents the electron-pair state being empty. Since Cooper pairs are formed by coherent superpositions of the ($\boldsymbol{k}_\uparrow, -\boldsymbol{k}_\downarrow$) electron-pairs with the ($\boldsymbol{k}_\downarrow, -\boldsymbol{k}_\uparrow$) hole-pairs on the Fermi surface, the BCS ground state can be described as a coherent superposition of the two pseudospin states, corresponding to the pseudospin lying within the xy-plane of the Anderson Bloch sphere (Fig. 7). With minimal coupling, the effect of an external electromagnetic field can be mapped to a pseudo-magnetic field pointed along the z-axis of the Bloch sphere, inducing a precession of the pseudospin at twice the electromagnetic driving frequency. Tsuji and Aoki simulated this $2\omega$ precession, from which a $3\omega$ component in the superconductor's current response is obtained [22].

This is the origin of the 2$\omega$ oscillation in the sample's transmissivity as well as the THG radiation from the sample. It is worth mentioning that in 2013, a near-infrared (1.55 eV) pump-while light (1.6 to 3.2 eV) probe study on the cuprate high-temperature superconductor La$_{2-x}$Sr$_x$CuO$_4$ reported free oscillations of 2$\Delta$ frequency around a specific wavelength of the probe light (~ 2 eV) after pump excitation [25]. In this work, such dynamics is interpreted as the quasiparticle dynamics near the Fermi surface, using also the Anderson pseudospin picture for an effective description. In retrospect, we cannot rule out that the quasiparticle dynamics reported in this work could in fact be the Higgs mode dynamics in La$_{2-x}$Sr$_x$CuO$_4$.

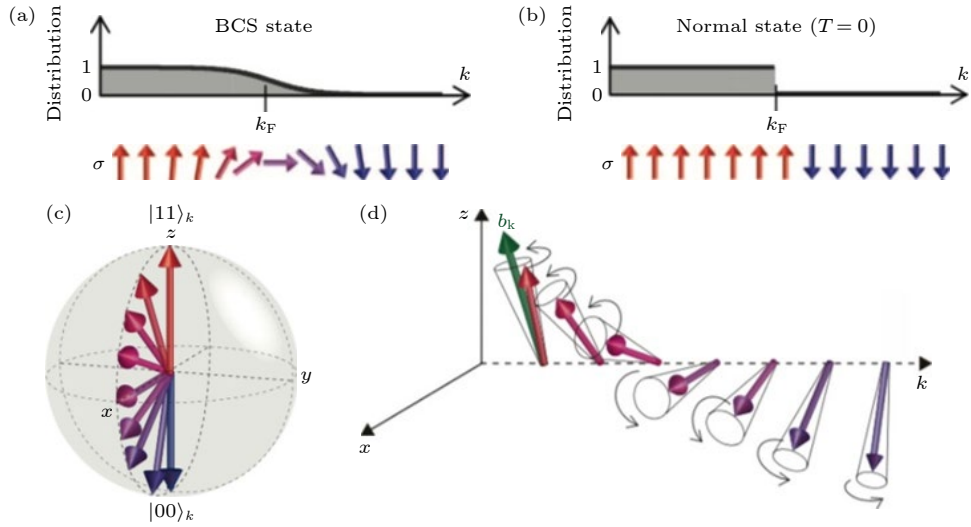

**Figure 7 Precession of the Anderson pseudospin under a multicycle terahertz drive** (reproduced with permission from Ref. [22]). **(a)(c)** In the BCS state, pairs of electrons at $(k_\uparrow, -k_\downarrow)$ and pairs of holes at $(k_\uparrow, -k_\downarrow)$ coherently superpose to form Cooper pairs. The occupancy of the pair of states at $(k_\uparrow, -k_\downarrow)$ can be used to define a pseudospin = 1/2 state (occupied) and a pseudospin = –1/2 state (empty). This way, the BCS ground state can be expressed as a superposition of the two pseudospin states, corresponding to a vector lying in the $xy$-plane of the Bloch sphere defined by the two pseudospin states. **(b)** In the normal metallic state, there is no coherence between the electrons at $(k_\uparrow, -k_\downarrow)$ and the holes at $(k_\uparrow, -k_\downarrow)$. Therefore, the Fermi surface is represented by a disrupt transition from the pseudospin = 1/2 state to the pseudospin = –1/2 state. **(d)** Under the periodic drive of a terahertz pulse, the Anderson pseudospin precesses. This corresponds to collective oscillations of the superconducting order parameter.

While the Anderson pseudospin provides a method for semi-classically describing the nonlinear current response of a superconductor under terahertz irradiation which in turn radiates THG, the same process can also be understood from a microscopic perspective of inelastic light-matter scattering (i.e. Raman scattering). The latter is particularly important for understanding why time-domain nonlinear terahertz spectroscopy can effectively couple to the Higgs mode, as well as the low-energy collective modes of other condensed phases in general. As mentioned earlier, the kinetic energy of an electron subject to an electromagnetic field is $[\mathbf{p} + e\mathbf{A}]^2/(2m)$, which can be expanded to obtain two expressions containing the electromagnetic vector potential: $[\mathbf{A} \cdot \mathbf{A}] \cdot e^2/(2m)$ and $[\mathbf{p} \cdot \mathbf{A} + \mathbf{A} \cdot \mathbf{p}] \cdot e/(2m)$. They represent two distinct types of electron-photon scattering in a solid. Assuming that the (many-body) electronic state of the solid before and after the scattering is $|I\rangle$ and $|F\rangle$ (note: $|I\rangle$ and $|F\rangle$ do not necessarily correspond to the eigenstates $|\alpha\rangle$, $|\beta\rangle$, ... of the system's Hamiltonian. They can be projected onto such eigenbasis as a linear combination of $|\alpha\rangle$, $|\beta\rangle$, ...). Using second quantization[§], the operator for the effective light-matter interaction $M$ can be expressed as [26]:

$$M_{I,F} = \mathbf{e}_i \cdot \mathbf{e}_s \sum_{\alpha,\beta} \rho_{\alpha,\beta}(\mathbf{q}) \langle F | c_\alpha^\dagger c_\beta | I \rangle$$

$$+ \frac{1}{m} \sum_v \sum_{\alpha,\alpha',\beta,\beta'} p_{\alpha,\alpha'}(\mathbf{q}_s) p_{\beta,\beta'}(\mathbf{q}_i) \times \left( \frac{\langle F | c_\alpha^\dagger c_{\alpha'} | v \rangle \langle v | c_\beta^\dagger c_{\beta'} | I \rangle}{E_I - E_v + \hbar\omega_s} + \frac{\langle F | c_\beta^\dagger c_{\beta'} | v \rangle \langle v | c_\alpha^\dagger c_{\alpha'} | I \rangle}{E_I - E_v - \hbar\omega_s} \right) \quad (6)$$

The first term on the right side of the equation, $\rho_{\alpha,\beta}(\mathbf{q}) \equiv \langle \alpha | e^{i\mathbf{q}\cdot\mathbf{r}} | \beta \rangle = \int d^3 r \varphi_\alpha^*(\mathbf{r}) e^{i\mathbf{q}\cdot\mathbf{r}} \varphi_\beta(\mathbf{r})$, is the matrix element of single-particle density fluctuation between the eigenstates of the Hamiltonian. It describes the probability amplitude of an inelastic scattering between an incoming photon and an electron, exchanging part of their energy and momentum (Fig. 8(a)), resulting in the system transiting from an initial state $|\alpha\rangle$ to a final state $|\beta\rangle$. A linear combination of these transition probability amplitudes (according to the projection of $|I\rangle$ and $|F\rangle$ onto the eigenbasis) then gives the probability amplitude of the transition from $|I\rangle$ to $|F\rangle$ by the above inelastic

---

[§] $c_\alpha^\dagger$ ($c_\alpha$) creates (annihilates) the state $|\alpha\rangle$. $\mathbf{e}_i$ ($\mathbf{e}_s$) denotes the polarization vector of the incident (scattered) photon. $\mathbf{q}_i$ ($\mathbf{q}_s$) denotes the momentum of the incident (scattered) light. In addition, we perform Fourier transform of the vector potential to obtain $\mathbf{A} = \sum_q e^{i\mathbf{q}\cdot\mathbf{r}} \mathbf{A}_q$, $\mathbf{A}_q = \mathbf{e}_q a_{-q} + \mathbf{e}_q^* a_q^\dagger$, where $a_q$ ($a_q^\dagger$) creates (annihilates) a photon.

scattering process. In the second term, $p_{\alpha,\alpha'}(\boldsymbol{q}_i) \equiv \langle \alpha | \boldsymbol{p} \cdot \boldsymbol{e}_i e^{i\boldsymbol{q}_i \cdot \boldsymbol{r}} | \alpha' \rangle$ is the momentum density matrix element. A linear combination of them (in the same sense as above) gives the probability amplitude that the incident photon is resonantly absorbed by an electron, promoting the system from the initial state $|I\rangle$ to an intermediate state $|v\rangle$. The probability amplitude of the conjugate (time-reversed) process, i.e. transition from the intermediate state $|v\rangle$ to the final state $|F\rangle$ through the radiation of a photon, is described by a linear combination of $p_{\beta,\beta'}(\boldsymbol{q}_s)$. The linear combination of the product of $p_{\alpha,\alpha'}(\boldsymbol{q}_i)$ with $p_{\beta,\beta'}(\boldsymbol{q}_s)$, summed over all intermediate states $|v\rangle$, gives the total probability amplitude from transitioning from $|I\rangle$ to $|F\rangle$ through the resonant scattering process described just now (Fig. 8(b)).

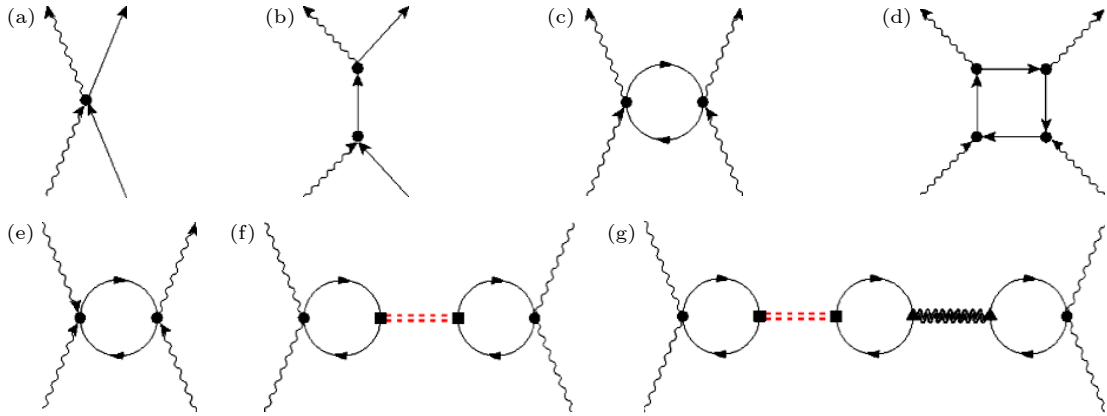

**Figure 8 Microscopic mechanisms for Raman scattering in solids** [26]. **(a)** Non-resonant photon-electron scattering (also known as two-photon scattering, intraband scattering, diamagnetic scattering) induces electron-hole density fluctuation. During the scattering, the energy and momentum of the photon and the electron both change. **(b)** In the resonant photon-electron scattering process (also known as interband scattering, paramagnetic scattering), a photon is absorbed by an electron and generates an electric current, which later re-radiates out a photon. **(c)** Microscopic process for non-resonant elastic (Rayleigh) scattering. **(d)** Microscopic process for resonant elastic scattering. **(e)** Microscopic process for terahertz third harmonic generation (THG). **(f)** Microscopic process for non-resonant Raman scattering or terahertz THG involving an intermediate boson. **(g)** Microscopic process for non-resonant Raman scattering or terahertz THG involving two intermediate bosons. Wavy lines represent photon propagators; solid lines represent electron propagators; red double dotted lines and black double wavy lines represent different boson propagators.

It is easy to recognize that the first term in $M$ corresponds to $[\mathbf{A} \cdot \mathbf{A}] \cdot e^2/(2m)$, while the second term corresponds to $[\mathbf{p} \cdot \mathbf{A} + \mathbf{A} \cdot \mathbf{p}] \cdot e/(2m)$ in the expansion of light-matter interaction. $M$ describes the scattering amplitude. The cross-section of light-matter scattering (i.e. the experimentally measured Raman scattering intensity) is proportional to $|M|^2$. Therefore, the cross-section of the two scattering mechanisms is given respectively by Fig. 8(c) & (d). It can be seen that both mechanisms involve two incoming photons and two outgoing photons. When we change an outgoing photon in Fig. 8(c) to an incoming photon in Fig. 8(e), we obtain the microscopic picture of terahertz THG. It should be pointed out that when higher-energy photons are used to produce THG in a semiconductor, the microscopic process may be described by Fig. 8(d) in cases where the photons resonantly excite interband electronic transitions. A careful reader may also notice that the scattering processes we described so far are all elastic. In these elastic scattering processes, we can insert an intermediate boson propagator (Fig. 8(f)) to obtain a description for inelastic Raman scattering. By analogy, from Fig. 8(c)(d) we may obtain the Feynman diagrams for non-resonant and resonant Raman scattering. Therefore, in solids the terahertz THG signal can be seen as the non-resonant Raman scattering signal of various collective modes.

Before we end this section, we elaborate on the terahertz THG process (Fig. 8(f)) in superconductors. In general, terahertz photons fall far below interband transitions in solids, therefore they mostly participate in inelastic intraband scattering with electrons. The cross-section of such process is identical to that of two-photon absorption (Fig. 8(a)), resulting in density fluctuation of an electron-hole pair. The latter may scatter a Higgs boson with energy $2\omega$ in a virtual process. The electron-hole pair finally recombines by scattering inelastically with a probe photon (of energy $\omega_{\text{probe}}$), resulting in an anti-Stokes shifted outgoing photon (of energy $\omega_{\text{probe}} + 2\omega$). This microscopic process explains the terahertz pump-terahertz probe experiments in Fig. 5 and Fig. 6, as well as terahertz pump-optical probe experiments [27,28]. In the terahertz THG experiment, although only a single terahertz beam is used, the multicycle terahertz pulse acts as both the pump and the probe. Therefore, it can also be regarded as a pump-probe technique in a general sense.

**Recent Progress**

After the pioneering works of Matsunaga and Shimano, ultrafast terahertz excitation followed by optical detection became a widely-adopted approach for studying the Higgs mode. However, Cea *et al.* questioned whether the third-order nonlinear signal $\chi^{(3)}(\omega_{\text{probe}} + 2\omega_{\text{THz}}, \omega_{\text{probe}}, \omega_{\text{THz}}, \omega_{\text{THz}})$ commonly observed in such experiments really arises from the Higgs mode or not [29]. In particular, they propose that single-particle density fluctuation (Fig. 8(e)) contributes more THG than the Higgs mode fluctuation (Fig. 8(f)). A caveat for distinguishing the two contributions is that the former gives rise to an anisotropic THG response consistent with the symmetry of the electronic band structure, while the latter yields an isotropic THG response. Matsunaga *et al.* examined the polarization dependence of THG from NbN and found a largely isotropic response, consistent a Higgs mode-origin of the THG signal [30]. Tsuji *et al.* pointed out that for BCS superconductors in the clean limit, the single-particle density fluctuation does contribute a stronger THG signal. However, most BCS superconductors are in the dirty limit. There, phonon retardation makes the Higgs mode-contribution dominate the THG response [31]. Cea et *al.* then pointed out that the terahertz pulse temporal width, the multiband nature of the electronic structure, as well as disorder in the system, might serve to erase the anisotropy in the THG signal originating from single-particle density fluctuations [32,33]. Subsequently, Tsuji and Nomura considered the different sources of THG in the clean and dirty limits of a multiband superconductor [34]. Through BCS self-consistent Born approximation, they find that the THG mediated by single-particle density fluctuation should still exhibit obvious anisotropy. On a separate note, Gabriele *et al.* pointed out that superconducting phase fluctuation (also known as Josephson fluctuation) also contributes to terahertz THG [35]. Moreover, this process is resonantly enhanced at $2\Delta(T) = 2\omega$ very similar to the Higgs mode contribution, making it difficult to assign the microscopic mechanism behind THG based solely on temperature dependence. It is worth noting that the above studies are generally based on mean-field superconductors. Whether the conclusions apply to strongly correlated systems or high-temperature superconductors is worth investigating. For example, in BCS superconductors, the Higgs mode is often rapidly damped by degenerate quasiparticle excitations. In a recent study, Lorenzana and Seibold found that strong correlation renormalizes the Higgs mode below the quasiparticle continuum, giving it a longer lifetime [36]. The coherent oscillation in the transient reflectivity of $La_{2-x}Sr_xCuO_4$ at $2\Delta$ frequency after optical excitation [25] may be taken as the evidence for a long-lived Higgs mode in high-temperature superconductors.

While the microscopic mechanism behind the nonlinear terahertz signal is being debated, Kemper *et al.* predicted the signature of the Higgs mode in time-resolved angle-resolved photoelectron spectroscopy (ARPES) experiment [37]. Krull *et al.* studied the coupling between the Higgs mode and the Legget mode in a multi-band superconductor such as $MgB_2$ [38]. Moor *et al.* proposed that the Higgs mode acquires infrared activity in the presence of a supercurrent [39], which was experimentally demonstrated by Nakamura *et al.* [40] Barlas and Varma predicted that *d*-wave superconductors manifest a collection of Higgs modes of different symmetries [41]. This is further elaborated in the work of Schwarz *et al.*[42]: applying a transient perturbation of different symmetries to *s*-wave, *d*-wave and *g*-wave superconductors leads to the generation of different Higgs modes with different symmetries. If so, the pairing symmetry of a superconductor may be revealed by counting the number of orthogonal Higgs modes found in a spectroscopy experiment. Yang *et al.* developed a gauge-invariant microscopic kinetic theory of superconductivity to study the response of superconductors to an optical field [43], and applied the theory to study the electromagnetic response of the superconducting phase mode and the Higgs mode [44]. Murakami *et al.* studied the coupling between the Higgs mode and phonons, which gives rise to new collective modes in addition to modifying the decay of the Higgs mode [45,46].

On the experimental side, Sherman *et al.* compared scanning tunneling spectroscopy to terahertz spectroscopy on disordered NbN superconductors, demonstrating the key difference between single-particle excitation and Higgs mode excitation [47]. Katsumi *et al.* used monocycle terahertz pulses as pump and 800 nm pulses as probe to study the Higgs response of $Bi_2Sr_2CaCu_2O_{8+x}$ high-temperature superconductors. Around time zero, 800 nm reflectivity is found to manifest a temporal modulation described by the modulus square of the terahertz electric field $E(t)$, which is consistent with the quadratic coupling between the superfluid and electromagnetic field [27]. In addition, the $|E(t)|^2$ response in the 800 nm reflectivity exhibits mostly an isotropic dependence on the angle between the terahertz field polarization and the 800 nm probe polarization. On top of the isotropic response, a small anisotropic component of $B_{1g}$ symmetry is found, indicating that the single-particle density fluctuation gives a small contribution to the third-order nonlinear optical response of the sample. In the multi-band superconductor $MgB_2$, Kovalev *et al.* used terahertz THG to study the Higgs response of the two coupled superconducting order parameters, but found evidence only for the collective response of the larger superconducting gap [48]. In NbN superconductors, Wang *et al.* employed moving wavelet analysis to study its nonlinear response to a multicycle terahertz field [49]. In addition to THG arising from the driven periodic oscillation of the superfluid, they also found evidence for a transient decay dynamics.

In 2020, Chu *et al.* reported the terahertz THG response of cuprate high-temperature superconductors [50]. In their analysis, they extracted the relative phase of the THG oscillation with respect to the linear periodic drive ($\Phi_{3\omega}$) and found that $\Phi_{3\omega}$ exhibits a sharp negative π jump with temperature (Fig. 9). By studying the phase evolution of a driven coupled oscillators model, they argued that the phase jump is a spectroscopic signature of anti-resonance, indicating the coupling of an overdamped Higgs mode with another underdamped collective mode in the system. Later, using the Fano resonance picture (i.e. interference between an underdamped excitation and an overdamped continuum) to describe the anti-resonance, they characterized the phase jump as a function of hole-doping and magnetic field in La$_{2-x}$Sr$_x$CuO$_4$ thin films. They observe that the phase jump in $\Phi_{3\omega}(T)$ is suppressed by overdoping and by magnetic field, which implies that the collective mode coupled to the Higgs mode could be CDW fluctuation [28] (Fig. 10). Building upon previous Raman scattering theory on 2*H*-NbSe$_2$ [52], Schwarz *et al.* considered the coupling of Higgs fluctuation to CDW fluctuation (Fig. 8(g)), and showed that terahertz THG would indeed manifest characteristics of an anti-resonance or Fano resonance with temperature due to such coupling [51]. To confirm this insight, Feng *et al.* characterized the THG response of a 2*H*-NbSe$_2$ thin film. They find that independent sources of THG originate below $T_{CDW}$ ~ 33 K and $T_c$ ~ 8 K respectively. The two THG signals strongly interfere near 6 K (Fig. 11), consistent with previous Raman scattering results on the sample (Fig. 3(b)) [53]. In the meantime, Yuan *et al.* also reported the beating of two independent THG signals near $T_c$ in the bilayer cuprate YBa$_2$Cu$_3$O$_{7-x}$ [54]. One of the THG signals originates at $T_c$ while the other persists to the pseudogap temperature $T^*$ (Fig. 12). They proposed that the latter may be contributed by a putative pseudogap collective mode. It is worth mentioning that the Cooper pair binding energy in cuprate superconductors falls generally in the range of 20-40 meV (i.e. 5-10 THz). In order to meet the resonance condition $2\Delta(T) = 2\omega$ of the Higgs mode, a driving frequency of $\omega$ = 2.5-5 THz should be used in experiment. However, this is beyond the spectral range of most table-top terahertz sources based on optical rectification in LiNbO$_3$. Nevertheless, due to the *d*-wave symmetry of the electron pairing in cuprate superconductors, the pairing energy 2Δ exhibits a continuous and periodic evolution around the Fermi surface. Therefore, the Higgs mode of *d*-wave superconductors is characterized by a broad spectral width. It is expected to manifest a strong spectral response even for driving frequency below $2\omega = 0.1 \times 2\Delta(T)$ [42]. This could be part of the reason why a terahertz drive between 0.3 and 0.7 THz still induces a significant THG response of cuprates in the above experiments. It may also justify why the Higgs mode of *d*-wave superconductors can be regarded as a continuum of excitations in the Fano resonance picture.

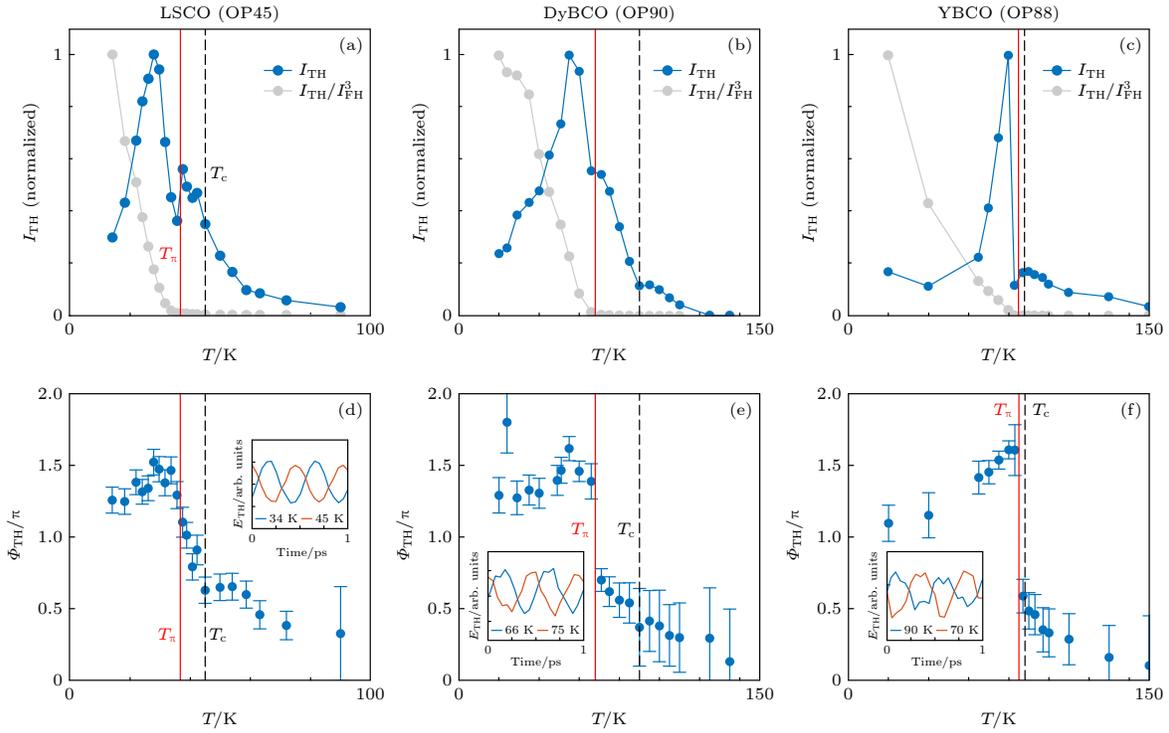

**Figure 9 Terahertz THG response of cuprate high-temperature superconductors** (reproduced with permission from Ref. [50]). **(a)-(c)** Temperature dependence of the terahertz THG intensity from optimally-doped $La_{2-x}Sr_xCuO_4$, $DyBa_2Cu_3O_{7-x}$, $YBa_2Cu_3O_{7-x}$ thin films. **(d)-(f)** Temperature dependence of the terahertz THG phase from the three samples. Black dotted line denotes $T_c$. Red solid line marks the temperature ($T_\pi$) at which the THG phase shows an abrupt jump of -π, which coincides with the local minimum in the THG amplitude response in (a)-(c). These features are well-described by the anti-resonance of a coupled oscillators model or the Fano resonance. Insets show the comparison of two representative time-domain THG responses below and above $T_\pi$.

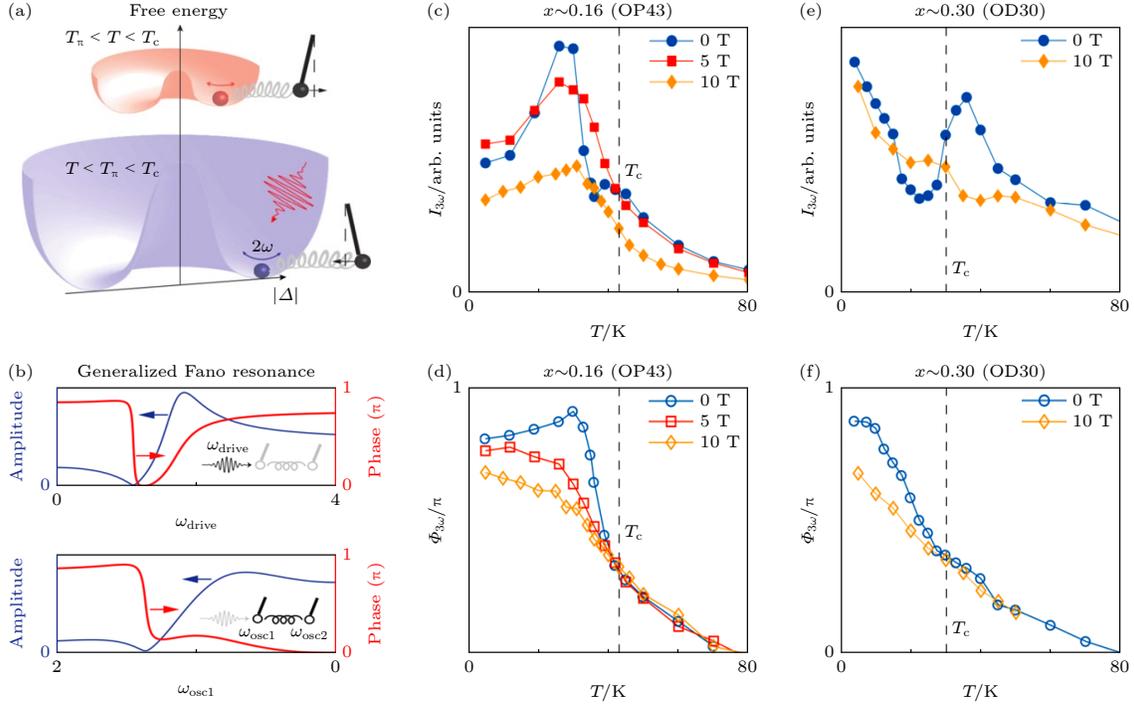

**Figure 10 Magnetic field dependence of the terahertz THG response of cuprate high-temperature superconductors** (reproduced with permission from Ref. [28]). **(a)** Illustration for the coupling of the Higgs mode to another collective mode. Below (above) $T_\pi$ the driven motions of the two oscillators are in-phase (out-of-phase), consistent with the phase evolution of the Fano resonance between an underdamped mode and a continuum background. **(b)** In typical spectroscopy manifestations, Fano resonance manifests as an asymmetric profile in the amplitude response and a negative jump in the phase response of the underdamped mode as a function of driving frequency ($\omega_{\text{drive}}$) or probe wavelength. In the terahertz THG experiment, the resonance frequency ($\omega_{\text{osc1}}$) of the collective mode is varied by varying temperature, while the driving frequency is kept constant. In this setting, the Fano resonance manifests as a similar evolution of the amplitude and phase response of the collective mode as function of $\omega_{\text{osc1}}$ or temperature. **(c)(e)** Temperature-dependent THG intensity from an optimally-doped and an overdoped $La_{2-x}Sr_xCuO_4$ thin film under different magnetic fields. **(d)(f)** Temperature-dependent THG phase of the two samples under different magnetic fields. The magnetic field is applied along the $c$-axis of the sample. It is clear that the $c$-axis magnetic field suppresses the Fano resonance in the THG response.

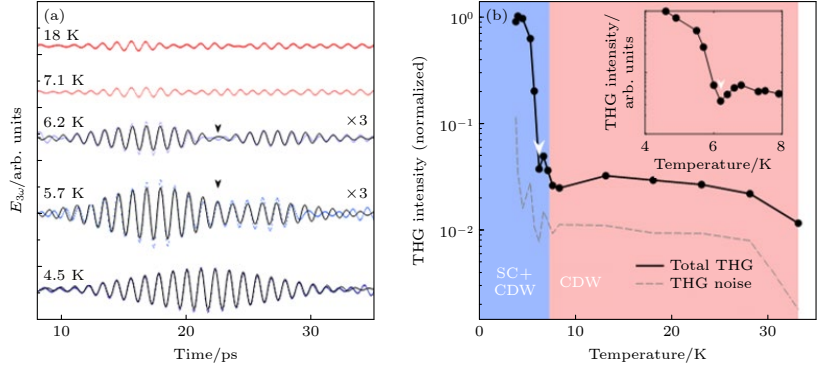

**Figure 11 Terahertz THG response of 2$H$-NbSe$_2$** (reproduced with permission from Ref. [53]). **(a)** Time-domain terahertz THG response of the sample in the superconducting and CDW states. Black arrow marks the interference between two distinct sources of THG. **(b)** Temperature dependence of the total terahertz THG intensity. Near 6 K, the destructive interference between the two sources of THG leads to a suppression of the total THG intensity. This in turn leads to a local minimum in the temperature dependence of the total THG intensity.

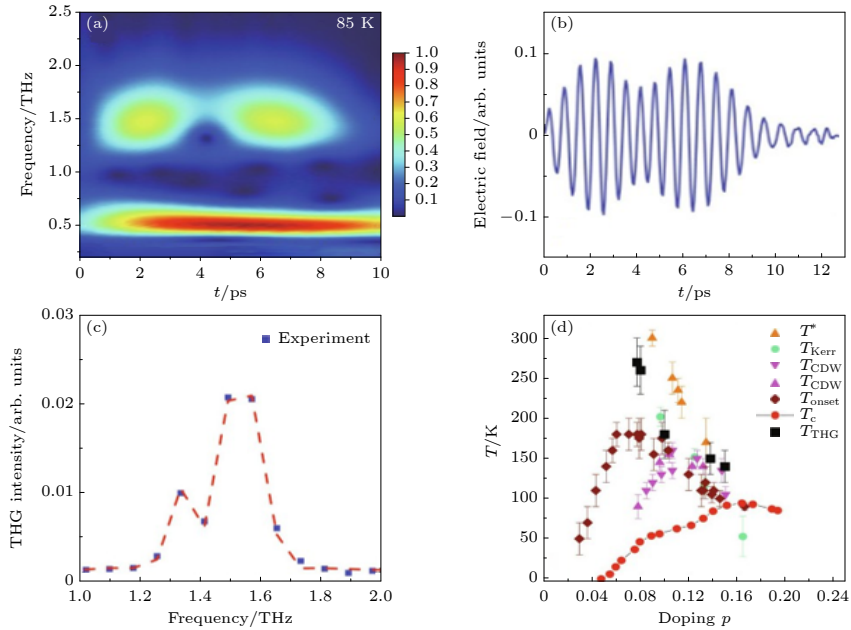

**Figure 12 Terahertz THG response of YBa$_2$Cu$_3$O$_{7-x}$** (reproduced with permission from Ref. [54]). **(a)** Moving-wavelet Fourier analysis of the terahertz transmission of an optimally-doped sample ($T_c$ = 87.4 K) under 0.5 THz periodic drive at 85 K. The THG response exhibits two individual wavelets separated in time. **(b)** Time-domain THG waveform. **(c)** Fourier analysis of the time-domain waveform shows two distinct components around the THG frequency. **(d)** The onset temperature for the terahertz THG response ($T_{THG}$) in the YBa$_2$Cu$_3$O$_{7-x}$ phase diagram. $T^*$ denotes the onset temperature for the pseudogap. $T_{Kerr}$ denotes the onset temperature for time-reversal symmetry breaking as reported by optical Kerr measurements. $T_{onset}$ denotes the onset temperature for superconducting fluctuations as determined by $c$-axis infrared ellipsometry measurements.

While nonlinear terahertz spectroscopy is being rapidly adopted for studying the Higgs mode, conventional resonant Raman scattering also made important progress in the field during the past decade. Using such technique, Méasson *et al.* systematically investigated the evolution of the Higgs mode and the CDW mode in layered materials such as $2H$ NbSe$_2$, $2H$ NbS$_2$, $2H$ TaS$_2$ as a function of temperature and pressure [55-57]. Glier *et al.* employed optical pump-Raman scattering probe to identify the anti-Stokes signal of the Higgs mode in Bi$_2$Sr$_2$CaCu$_2$O$_{8+\delta}$ (Bi-2212) superconductor after resonant excitation [58], in an effort to validate the theoretical prediction by Puviani *et al.*: the Higgs mode contributes a significant amount of spectral weight under the Raman scattering peak at $2\Delta$ [59], which has been long interpreted as entirely due to pair-breaking.

**Look to the Future**

As showcased above, nonlinear terahertz scattering and resonant Raman scattering have advanced the study of Higgs modes in various superconductors. An understanding of the Higgs mode dynamics in the perturbatively-driven regime is expected to bring new insights on equilibrium superconductivity, particularly for unconventional superconductors. It may provide fresh perspectives on the microscopic mechanism behind the pseudogap and the charge order, and the competition/co-operation between CDW and superconductivity. Previous studies on these intertwined orders have focused on their static or equilibrium characterization, such as the ordering temperature or the correlation length, under different physical conditions. For example, in many materials the application of a magnetic field suppresses superconductivity and enhances the charge order correlation, for example, by increasing its ordering temperature. From such observation, it is tempting to conclude that the two orders compete with each other. To describe such interplay, the coupled Ginzburg-Landau model is typically used:

$$F = a|\psi_{\text{SC}}|^2 + b|\psi_{\text{SC}}|^4 + \alpha|\phi_{\text{CDW}}|^2 + \beta|\phi_{\text{CDW}}|^4 + \lambda|\psi_{\text{SC}}|^2|\phi_{\text{CDW}}|^2 \qquad (7)$$

which is equivalent to a coupled oscillators model. The ordering of different order parameters in thermal equilibrium is captured by the mean-field solution of the free energy minimization problem, or the static displacements of the coupled pendula (Fig. 13(a)). By observing a correspondence between the two pendula's static displacements, one may be brought awareness of the coupling between the pendula, but still learns nothing about the energy scale nor the microscopic mechanism behind the coupling. Only when the coupled system is dynamically perturbed can one obtain such microscopic information (Fig. 13(b)). In the dynamical picture, whether two orders or two pendula co-operatively interact (i.e. their fluctuations are in-phase) or competingly interact (i.e. their fluctuations are out-of-phase) becomes a subtle question: it depends on which frequency range of the fluctuations or the dynamics we are looking at. These are the unique perspectives that the Higgs mode, and more generally the dynamics of superconductors, may provide for understanding the superconductivity-CDW interplay.

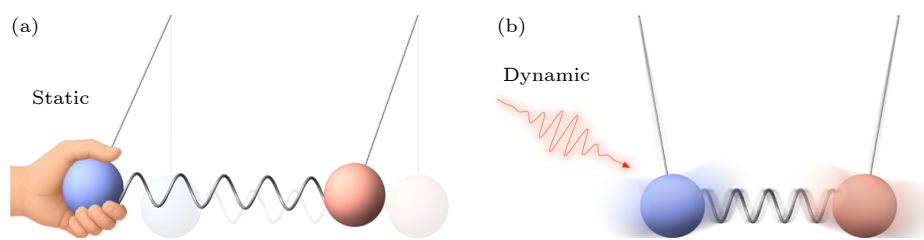

**Figure 13 Experimental approaches for investigating intertwined orders: equilibrium versus non-equilibrium. (a)** For investigating the interplay between intertwined orders (e.g. superconductivity and CDW), past studies have focused on their equilibrium characterizations. For example, by using an external magnetic field or strain field to suppress superconductivity, the resulting changes to the CDW correlation can be probed using x-ray diffraction. While such an approach provides evidence for the interplay between the two orders and allows it to be characterized as competing or co-operative in the static sense, it does not provide microscopic information about the interaction mechanism. This equilibrium strategy can be compared to displacing a coupled pendulum. By observing the static displacements of the two pendula, one may infer that they are coupled. However, no more detail about the coupling spring can be learnt using such an approach. **(b)** When the coupled pendula are dynamically perturbed, the energy and momentum transfer between the two pendula depends sensitively on the spring constant. Therefore, more specific details about the 'spring', such as its energy scale and microscopic mechanism, can be learnt from the dynamical response of the system.

As mentioned above, in the Standard Model of particle physics, in addition to mass generation for $W\pm$, $Z$ bosons via the Higgs mechanism, through its own spontaneous symmetry-breaking the Higgs field also renormalizes many fermionic fields through the Yukawa interaction. It could be interesting to search for analogues of the Yukawa interaction in condensed matter systems. For example, in addition to the specific form of superconductivity-CDW interaction discussed above, one could ask if there is any non-biquadratic coupling (i.e. of a form different from

$\lambda|\psi_{SC}|^2|\phi_{CDW}|^2$ in Eqn. (7)) between the different order parameters in a superconductor, such that the condensation (i.e. spontaneous symmetry-breaking) of one order renormalizes another co-existing order. For example, in the heavy fermion superconductor URu$_2$Si$_2$ [60], the spin liquid candidate system Cd$_2$Re$_2$O$_7$ [61], and iron-based superconductors [62], an avalanche of symmetry-breaking transitions takes place one after another in a narrow region of the phase diagram, possibly due to the coupling between different order parameters, which is not necessarily biquadratic. These materials provide an interesting platform for searching for the analogue of Yukawa interaction in solids. In some of these systems, the exact form of interaction (e.g. whether $\lambda\psi\phi^2$ or $\lambda\psi^2\phi^2$) is unveiled by symmetry analysis of the material's transport response under different symmetry-breaking fields (e.g. magnetic field, strain field) [61,62]. In comparison, the dynamical approach illustrated in Fig. 13(b) has a similar spirit as particle physics experiment: excitation of various collective modes from their ground states corresponds to the preparation of high energy particles before collision. The subsequent interaction between the collective modes can be seen as the scattering/collision events. The practicality of such 'particle-physics' approach is being demonstrated presently in the study of superconductivity-CDW interplay in Ref. [28,53,54] and in principle could be applied for studying a broader class of intertwined orders.

To fully enable these investigations, novel spectroscopy techniques are also urgently anticipated. For one example, terahertz two-dimensional coherence spectroscopy (2DCS) is a very relevant technique for investigating the coupling between different collective modes or degrees of freedom (Fig. 14) [63]. Visible and infrared 2DCS has long been employed for studying excitons in semiconductors and molecular dynamics in physical chemistry research. Its application in the terahertz wavelength regime and to condensed matter systems is seeing rapid development in recent years. Using such technique, Vaswani *et al.* studied the hybridization between the Higgs modes of different superconducting order parameters in an iron-based superconductor [64]. Also in this system, Luo *et al.* further distinguished the different sources of high-order nonlinear optical signals, including the Higgs mode, bi-Higgs mode, etc. [65] Using a similar technique, Katsumi and Armitage reported a $\chi^{(3)}(\omega, \omega, -\omega, \omega)$ response in addition to the $\chi^{(3)}(3\omega, \omega, \omega, \omega)$ response from NbN superconductors, and argued that the $\chi^{(3)} \omega$ response provides a more unambiguous signature of the Higgs mode than the $\chi^{(3)}$ THG response [66]. Kim *et al.* used optical pump-terahertz THG probe to study the dynamics of the Higgs mode after 1.55 eV pump excitation in a cuprate superconductor [67]. An important caveat of 2DCS is that the excitation pulse has wide spectral width to cover multiple excitations. Through the coherence between the spectral components of the excitation pulse, coherence between the excited states (i.e. from a shared ground state) is established. A second and a third pulse manipulate the subsequent evolution of the coherent state, from which the coupling between different excited states and their decoherence processes can be revealed.

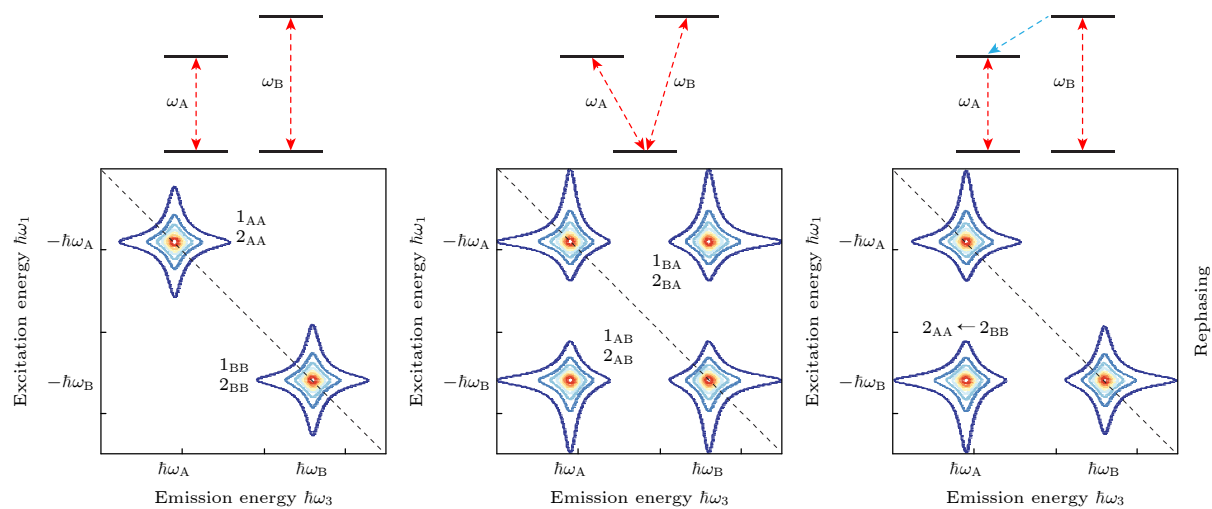

**Figure 14 Two-dimensional coherence spectroscopy (2DCS)** (reproduced with permission from Ref. [63]). In molecular and solid state systems, the coupling between different degrees of freedom (DOFs) and their excitations is often the key to understanding their physical or chemical properties or processes. **(a)** While conventional linear spectroscopy techniques (e.g. absorption, reflection, Raman scattering) are good at capturing the excitations of individual DOFs, evidencing their coupling is often challenging. **(b)(c)** In this regard, 2DCS has demonstrated significant potential. 2DCS uses several broadband pulses to create and manipulate coherence between excitations. The first pulse excites multiple excitations. (b) If the excitations share the same ground state, or (c) if they have different ground states but their excited states are coupled, then coherence can be established between the excited states by the excitation pulse. After a variable time-delay τ, a second and third pulse interact with the sample, inducing an evolution of the coherent state (e.g. rephasing or non-rephasing). Finally, the coherent state radiates and the system falls back to the ground state. By measuring the radiation spectrum for each τ, and Fourier-transforming it with respect to τ, a 2DCS map as illustrated above is obtained. The peaks lying on the diagonal of the map represents individual excitations. The peaks lying on the off-diagonal represents coherence between two excitations. Compared to traditional linear spectroscopy techniques, 2D spectroscopy provides more direct and detailed information about the interaction between different DOFs [63].

With the expansion of the spectroscopy toolbox, novel superconducting samples and devices will further enrich and advance the study of Higgs mode. The fabrication of ultra-thin samples with precisely controlled atomic layers by mechanical exfoliation or molecular beam epitaxy may promote a confluence of Higgs mode research and low-dimension superconductivity research. This is in fact an interesting area to test the validity of Anderson's mechanism. Mermin-Wagner theorem predicts that two-dimensional systems with continuous symmetry cannot sustain long-range ordering at temperatures above zero, due to enhanced thermal fluctuations of the gapless Goldstone mode in low dimension. Therefore, when considering a putative two-dimensional superconductor, it is critical but unclear which of the two theorems, Anderson's mechanism and the Mermin-Wagner theorem, takes precedence. If Anderson's mechanism takes precedence, it seems to suggest that pure two-dimensional superconductivity is possible. Otherwise, Mermin-Wagner theorem forbids pure two-dimensional superconductivity. Such ambiguity can perhaps be tested and further studied on many van der Waals superconducting systems readily available nowadays. Along this train of thought, Higgs mode research may also bring novel perspective on the Berezinskii-Kosterlitz-Thouless transition in 2D superconductivity research.

At present, many unknowns remain also in the spectroscopic response of superconducting devices of meso- and nanoscopic scale. In 2024, Du *et al.* reported Raman scattering investigation of the Higgs mode in a micrometer-sized bilayer $NbSe_2$ thin film in the presence of a magnetic field, where they related the Higgs mode response to the anomalous metal state that emerges at high magnetic field [68]. This study demonstrates for a first time how Higgs spectroscopy and transport techniques may complement each other in the study of superconductivity and other correlated phases. We envision that studies in the future will further look into the dynamical response of the unconventional superconducting state in twisted bilayer graphene, the Fulde-Ferrell-Larkin-Ovchinnikov (FFLO) state in transition metal dichalcogenide under a magnetic field, as well as Josephson junctions. At present, these systems mainly rely on transport techniques for characterization, which probe the ground state. A more complete physical picture will emerge by also understanding the dynamics of these systems.

Finally, Higgs mode is also an integral part of non-equilibrium and photo-induced superconductivity research. Recently, using nonlinear terahertz response as a measure for the superconducting order parameter, Isoyama and Katsumi *et al.* studied the photo-induced enhancement of superconductivity in iron and cuprate superconductors pumped by 1.55 eV and mid-infrared light respectively [69,70]. Collado *et al.* proposed that in periodically-driven superconductors, a population inversion of the quasiparticle distribution can be realized, giving rise to a novel Rabi-Higgs mode [71]. When dissipation is also considered, they found that the superconductor will exhibit distinct dynamical phases including chaos and the time crystal state as a function of the driving amplitude, frequency, and dissipation rate [72,73]. Fan *et al.* investigated the spatial distribution of the time crystal phase in periodically-driven superconductors [74]. If these predictions can be successfully demonstrated, we will have a more complete and fundamental toolbox for engineering superconducting devices and manipulating superconductivity down to microscopic length scale and ultrafast time scale. A better understanding of the dynamical response of superconductors, in either the perturbatively-driven limit or the non-perturbatively-driven limit, promises many exciting discoveries to arrive.